# Suppression of Antiferromagnetic Order by Strain in Honeycomb Cobaltate: Implication for Quantum Spin Liquid

Gye-Hyeon Kim,[1,*] Miju Park,[1,*] Uksam Choi,[1] Baekjune Kang,[1] Uihyeon Seo,[1] GwangCheol Ji,[2] Seunghyeon Noh,[3] Deok-Yong Cho,[4] Jung-Woo Yoo,[3] Jong Mok Ok,[2] and Changhee Sohn[1,†]

[1]Department of Physics, Ulsan National Institute of Science and Technology, Ulsan, 44919, Republic of Korea

[2]Department of Physics, Pusan National University, Pusan, 46241, Republic of Korea

[3]Department of Materials Science and Engineering, Ulsan National Institute of Science and Technology, Ulsan, 44919, Republic of Korea

[4]Department of Physics, Jeonbuk National University, Jeonju, 54896, Republic of Korea

**Abstract**

Recently, layered honeycomb cobaltates have been predicted as a new promising system for realizing the Kitaev quantum spin liquid, a many-body quantum entangled ground state characterized by fractional excitations. However, these cobaltates, similar to other candidate materials, exhibit classical antiferromagnetic ordering at low temperatures, which impedes the formation of the expected quantum state. Here, we demonstrate that the control of the trigonal crystal field of Co ions is crucial to suppress classical antiferromagnetic ordering and to locate its ground state in closer vicinity to quantum spin liquid in layered honeycomb cobaltates. By utilizing heterostructure engineering on $Cu_3Co_2SbO_6$ thin films, we adjust the trigonal distortion of $CoO_6$ octahedra and the associated trigonal crystal field. The original Néel temperature of 16 K in bulk $Cu_3Co_2SbO_6$ decreases (increases) to 7.8 K (22.7 K) in strained $Cu_3Co_2SbO_6$ films by decreasing (increasing) the magnitude of the trigonal crystal fields. Our experimental finding substantiates the potential of layered honeycomb cobaltate heterostructures and strain engineering to accomplish the extremely elusive quantum phase of matter.

# I. INTRODUCTION

The small amount of broken local symmetry has been blamed for making a whole many-body quantum entangled state collapse into a (quasi)classical product state. Ever since P. W. Anderson's idea of resonating valence bond [1,2], searching for many-body quantum entangled ground states and low-energy excitation in the matter has remained an important, but highly elusive task in condensed matter physics [3-5]. An outstanding example is the Kitaev quantum spin liquid (QSL), in which bond-dependent Ising interactions of spins on a two-dimensional honeycomb lattice induce macroscopic quantum entanglement and fractional excitations [6]. While intensive studies have found signatures of fractionalization in candidate materials such as $Na_2IrO_3$ [7-11] and α-$RuCl_3$ [12-18], all of them have shown classical long-range antiferromagnetic orderings at sufficiently low temperatures ($T$). The convincing hypotheses about the collapse of the predicted quantum state in low $T$ have pointed out broken local symmetry as a culprit [10,18]. For example, finite trigonal distortions act as one of the main effects in generating non-Kitaev spin interactions in various candidate materials [18-20], deviating from the original concept of realizing Kitaev QSL through edge-shared cubic octahedra as shown in Fig. 1(a) [21]. Therefore, experimental verification of strong correlations between local distortions and stability of classical ground states would mark a significant milestone in realizing these notoriously elusive quantum phases of matter.

In this context, $Cu_3Co_2SbO_6$, a layered honeycomb cobaltate, can be an exemplary model system for probing the hypothesized relationship between local trigonal distortion and magnetic ground state. Recent theoretical [19,22-24] and experimental [25-29] studies suggested layered honeycomb cobaltate can be in close vicinity of Kitaev QSL as their orbital structures can be described as spin-orbit entangled $J_{eff}$ = 1/2 states [23]. However, due to the relatively small spin-orbit coupling strength of Co ions, the presence of compressive trigonal distortion is expected to break $J_{eff}$ = 1/2 orbital pictures rather easily, favoring the stabilization of a single $a_{1g}$ and doubly degenerate $e_g$ orbitals [Fig. 1(a)] [30]. Therefore, previous theoretical studies suggested that the Kitaev QSL state could be realized only with a small trigonal crystal field ($\Delta_{trig}$) in the layered honeycomb cobaltates [23]. As shown in Fig. 1(b), $Cu_3Co_2SbO_6$ has distorted local octahedra and displays classical antiferromagnetic ordering near 16 K [31] consistent with the aforementioned theoretical hypotheses. Notably, unlike other Kitaev material candidates, this compound has been successfully synthesized in thin film [32], facilitating the application of lattice engineering, which enables manipulation of local distortion of $CoO_6$ octahedra [33-35].

Here, we present experimental evidence that the control of the local $\Delta_{trig}$ induces a wide modulation in the Néel temperature ($T_N$) of $Cu_3Co_2SbO_6$. By utilizing $Cu_3Co_2SbO_6$ (001) film grown on ZnO (0001) substrate [32], we successfully manipulated the trigonal distortion of the $CoO_6$ octahedra and $\Delta_{trig}$ of Co ions [Fig. 1(c)]. We observed that the ultrathin $Cu_3Co_2SbO_6$ film got strained on the ZnO substrate, notwithstanding the large lattice mismatch (+ 4.3 %). This strain allowed us to stretch the octahedra in the in-plane direction, resulting in an increased compressive trigonal distortion. Due to the absence of a suitable substrate for out-of-plane stretching in the octahedra, we conducted helium implantation on the bulk-like 20-unit cell (u.c.) $Cu_3Co_2SbO_6$ films. This is a well-known methodology for out-of-plane expansion of $CoO_6$ octahedra thereby mitigating the compressive trigonal distortion [35]. X-ray absorption spectroscopy (XAS) reveals a clear increase (decrease) of $\Delta_{trig}$ of $CoO_6$ octahedra by reducing film thickness (by helium implantation). Consequently, the original $T_N$ of 16 K in $Cu_3Co_2SbO_6$ was modulated from 22.7 K to 7.8 K, depending on the direction and the magnitude of the applied stretching. The modulation of $T_N$ is as large as 90 % of the original $T_N$, indicating a strong correlation between $\Delta_{trig}$ and the stability of classical antiferromagnetic ground states. Our findings imply the possibility of a genuine spin liquid state if more stretching along the out-of-plane direction can be applied on $Cu_3Co_2SbO_6$.

## II. ORBITAL STRUCTURE OF $Co^{2+}$ ION IN HONEYCOMB COBALTATES

We first compare Co $L$-edge XAS of bulk-like 20 u.c. $Cu_3Co_2SbO_6$ film with that of $Na_3Co_2SbO_6$ film to elucidate the rather controversial relationship between orbital structures and distorted local $CoO_6$ octahedra in layered honeycomb cobaltates. Note that $Na_3Co_2SbO_6$ has an almost identical local structure of $CoO_6$ octahedra to $Cu_3Co_2SbO_6$, except for its smaller trigonal distortion (Fig. S1 in Ref. [36]). Considering crystal fields only from $O^{2-}$ ions, we can expect positive $\Delta_{trig}$ with a lower energy of $a_{1g}$ orbitals than $e_g^\pi$ orbitals in both compounds (Fig. S2 in Ref. [36]), as the octahedra are compressed in the out-of-plane direction [Fig. 1(a)]. However, previous theoretical and experimental research on $Na_3Co_2SbO_6$ suggested the negative value of $\Delta_{trig}$ [23,37]. It has been attributed to that the size of the negative crystal field from $Sb^{5+}$ ions is larger than the size of the positive crystal field from $O^{2-}$ ions (Fig. S2 in Ref. [36]).

Despite those predictions for $\Delta_{trig}$, the $Cu_3Co_2SbO_6$ displayed a notably larger X-ray linear dichroism (XLD) in Co $L_3$-edge spectra than $Na_3Co_2SbO_6$, implying a more dominant role of $O^{2-}$ than $Sb^{5+}$ ions in $\Delta_{trig}$ and the positive sign in total $\Delta_{trig}$. Figures 2(a) and 2(b) exhibit the normalized isotropic XAS, $I_{iso} = (2I_{\perp z} + I_{\|z}) / 3$, and XLD, $I_{XLD} = I_{\perp z} - I_{\|z}$, at the Co $L_3$ edges for $Cu_3Co_2SbO_6$ and $Na_3Co_2SbO_6$, respectively. Here, $I_{\perp z}$ ($I_{\|z}$) denotes the XAS spectra with the incident polarization (denoted as **E**) of light perpendicular (parallel) to the out-of-plane direction. We changed the polarization of the incident X-ray with a fixed incident angle $\alpha$ of 70° (Fig. S3 in Ref. [36]). From this experimental geometry, the spectra for the polarization were obtained by using the formula, $I_{\perp z} = I_p$ and $I_{\|z} = [I_s - I_p\cos^2(\alpha)] / \sin^2(\alpha)$, where $I_p$ ($I_s$) are the measured intensities with $p$ ($s$) polarization. The $I_{iso}$ spectra at the Co $L_3$ edge of both compounds are nearly identical and resemble the well-known spectra of $Co^{2+}$ ions [38]. The major difference in those spectra is that the $I_{XLD}$ of $Cu_3Co_2SbO_6$ is roughly two times larger than that of $Na_3Co_2SbO_6$, indicating a larger magnitude of $\Delta_{trig}$ [39]. This result contrasts with the previous theoretical predictions which suggested a dominant $Sb^{5+}$ crystal field and a negative $\Delta_{trig}$. If $\Delta_{trig}$ were negative in layered honeycomb cobaltates, a more distortion in octahedra in $Cu_3Co_2SbO_6$ than $Na_3Co_2SbO_6$ would reduce the overall magnitude of $\Delta_{trig}$ with a larger positive crystal field from $O^{2-}$. The comparison between these two similar compounds underlines the dominant influence of the $O^{2-}$ crystal field and supports a positive $\Delta_{trig}$ in layered honeycomb cobaltates (Fig. S2 in Ref. [36]). For estimating the magnitude of $\Delta_{trig}$, the crystal field multiplet calculations were conducted using the quantum many-body script language, Quanty [40] with Crispy interface [41]. Even in the multiplet calculations, simulations with $\Delta_{trig}$ = 37.8 meV (25.1 meV) are good agreements with the experimental $I_{XLD}$ of $Cu_3Co_2SbO_6$ ($Na_3Co_2SbO_6$) film [Figs. 2(c) and 2(d)] and assist the above qualitative analysis.

The polarization-dependent O $K$-edge spectra further clarify unoccupied $e_g^{\pi}$ orbitals and positive $\Delta_{trig}$ in the layered honeycomb cobaltates. The $Co^{2+}$ ions in layered honeycomb cobaltates exhibit high spin $d^7$ configurations. Therefore, in O $K$-edge XAS of $Cu_3Co_2SbO_6$ and $Na_3Co_2SbO_6$ [Figs. 2(d)-2(f)], the lowest energy peak near 531 eV can be assigned to O $2p$ - Co $t_{2g}$ hybridized state, while the multiple-peak features at higher energies are associated with O $2p$ orbitals hybridized with unoccupied Co $e_g^{\sigma}$ and Cu $3d$ orbitals (Fig. S4 in Ref. [36]) [42-

46]. Given that trigonal distortion splits only the $t_{2g}$ orbitals, we focused on polarization dependence in the lowest peak. In the O $K$ edge, $I_{\perp z}$ ($I_{\parallel z}$) of the peak is proportional to the averaged interatomic matrix element $V^2_{pd}$ between Co $t_{2g}$ orbitals and O $2p$ orbitals perpendicular (parallel) to the out-of-plane direction. Consequently, the relative intensity between $I_{\perp z}$ and $I_{\parallel z}$ is sensitive to whether the orbital character of the unoccupied $t_{2g}$ orbital is mainly $a_{1g}$ or $e_g^\pi$. The intensity ratio, $I_{\perp z} / I_{\parallel z}$, for the $a_{1g}$ orbital character is calculated to be 0.25 in undistorted cubic octahedra and is expected to be less than 0.25 in the presence of trigonal distortions. That for $e_g^\pi$ symmetry, on the other hand, is expected to be 1 in undistorted cubic octahedra and larger than 1 with trigonal distortion [43]. Therefore, if those cobaltates had a negative $\Delta_{trig}$ and unoccupied $a_{1g}$ orbitals as predicted, we would observe a strong suppression in Co $t_{2g}$ peak intensity in $I_{\perp z}$ compared to $I_{\parallel z}$ spectra. However, both compounds show enhancement of the peak intensity in $I_{\perp z}$ than in $I_{\parallel z}$, unambiguously indicating that the unoccupied $e_g^\pi$ orbitals are dominant and that $\Delta_{trig}$ is indeed positive.

### III. STRAIN ENGINEERING IN $Cu_3Co_2SbO_6$ HETEROSTRUCTURE

By utilizing our heterostructure geometry, we have successfully controlled the crystal structure and $\Delta_{trig}$ of $Cu_3Co_2SbO_6$ thin films. Figure 3(a) exhibits X-ray diffraction (XRD) $\theta$-$2\theta$ scan near the (008) peak of $Cu_3Co_2SbO_6$ film with 20 u.c. and 3 u.c. thickness. The (008) peak of 3 u.c. $Cu_3Co_2SbO_6$ film shifts to a higher angle than that of 20 u.c. $Cu_3Co_2SbO_6$ film, indicating reduced out-of-plane lattice constant, $c^*$ ($c^* = c\sin(\beta)$ in lattice parameter) with thickness [33]. As shown in Fig. 3(b), $c^*$ is quickly relaxed to its bulk value with increasing thickness due to the large lattice mismatch (+ 4.3 %) between ZnO and $Cu_3Co_2SbO_6$ (Fig. S5 in Ref. [36]). Reciprocal space mapping (RSM) data and $q_x$ profiles [Figs. 3(c) and 3(d)] of the $Cu_3Co_2SbO_6$ (606) plane further demonstrates that the $q_x$ value for the 3 u.c. film is smaller than that of the 20 u.c. film, indicative of an in-plane elongation. Based on these findings, we conclude that, despite the significant lattice mismatch, a tensile strain is indeed imparted upon the ultrathin $Cu_3Co_2SbO_6$ films. To increase $c^*$, on the other hand, we adopted helium implantation as helium atoms in matter induce $c^*$ expansion with epitaxially locked in-plane lattice constants [34,35]. As we increased the dose of helium ions, the (004) peak of

$Cu_3Co_2SbO_6$ films shifted to a lower angle, in consistent with increased $c^*$ [Figs. 3(e) and 3(f)]. Same as thickness-controlled films, we obtained RSM and $q_x$ profiles of the $Cu_3Co_2SbO_6$ (606) plane for the He $3.6 \times 10^{15}$ cm$^{-2}$ implanted film [Figs. 3(g) and 3(h)]. As we expected, helium implantation barely modulates in-plane lattice constant compared to pristine $Cu_3Co_2SbO_6$ due to its locked lattice by neighboring unit cells. Full RSM images for each sample are shown in Fig. S6 in Ref [36].

The observed decrease (increase) of $c^*$ in $Cu_3Co_2SbO_6$ films indeed reinforces (undermines) the trigonal distortion of the octahedra and $\Delta_{trig}$. Figure 4(a) exhibits the evolution of $I_{XLD}$ of Co $L_3$ edge with thickness and He ion implantation. $I_{XLD}$ of 3 u.c. $Cu_3Co_2SbO_6$ film shows an overall enhancement compared to the 20 u.c. $Cu_3Co_2SbO_6$ film, indicating an increased $\Delta_{trig}$ [Fig. 4(a), left]. The He $3.6 \times 10^{15}$ cm$^{-2}$ implanted $Cu_3Co_2SbO_6$ films, in contrast, exhibit smaller $I_{XLD}$ relative to the pristine $Cu_3Co_2SbO_6$ film [Fig. 4(a), right], indicating a decreased $\Delta_{trig}$. Figure 4(b) shows the absolute area of XLD, $I_{|XLD|}$, in each stretched-thin film, which roughly represents the relative magnitude of $\Delta_{trig}$ [39], as well as simulated $\Delta_{trig}$, which reproduced experimental $I_{XLD}$ (Fig. S7 in Ref. [36]). Notably, $I_{|XLD|}$ of both the 3 u.c. and the He $3.6 \times 10^{15}$ cm$^{-2}$ implanted film is increased (reduced) to approximately 13 % (33 %) of the original value in the bulk-like $Cu_3Co_2SbO_6$, demonstrating capabilities of heterostructures for the direct control of Hamiltonian parameters.

## IV. MODULATION OF THE NEEL TEMPERATURE INDUCED BY STRAIN ENGINEERING

To investigate the effect of the $\Delta_{trig}$ on its magnetic ground state, we conducted spectroscopic ellipsometry to detect $T_N$ in strained $Cu_3Co_2SbO_6$ films. The previous research on $Cu_3Co_2SbO_6$ has shown a peculiar spin-exciton coupling that induces a distinct kink at $T_N$ in a raw ellipsometric parameter $\Psi$, where $\tan\Psi$ is the intensity ratio between reflected $p$- and $s$-polarized light at the exciton peak energy of ~ 4 eV (Fig. S8 in Ref. [36]). Therefore, we fixed the photon energy at the resonant frequency of the exciton and obtained temperature-dependent $\Psi$ as shown in Figs. 5(a)-5(f). Note that conventional magnetometry experiments make it difficult to detect $T_N$ of ultrathin film owing to its extremely small volume and large paramagnetic/diamagnetic backgrounds from substrates, defects, and equipment environment (Fig. S9 in Ref. [36]). This optical detection of $T_N$ is not only sufficiently sensitive for ultrathin films but also free from any paramagnetic and diamagnetic backgrounds as no external magnetic field is required. Figure 5(a) shows Ψ (*T*) of 20 u.c. $Cu_3Co_2SbO_6$ thin film. Even without complicated model fitting, we observed a clear kink at ~ 16 K in the raw ellipsometry parameter, which is consistent with $T_N$ observed in conventional magnetometry experiments.

The strain engineering on $Cu_3Co_2SbO_6$ films and the resultant change in $\Delta_{trig}$ induced a massive modulation in $T_N$. Figures 5(a)-5(f) exhibit optical parameter Ψ of strained $Cu_3Co_2SbO_6$ thin films as a function of *T*. While $T_N$ of the 3 u.c. $Cu_3Co_2SbO_6$ film with a larger $\Delta_{trig}$ is shifted to 22.7 K ± 1.8 K [Fig. 5(c)], $T_N$ in the He 3.6 × $10^{15}$ $cm^{-2}$ $Cu_3Co_2SbO_6$ films with a reduced $\Delta_{trig}$ is suppressed to 7.8 K ± 1 K [Fig. 5(f)]. The reduced $T_N$ in helium-implanted films can hardly be attributed to any extrinsic origins such as impurity formation, as it recovered to the original value of 16 K with the removal of the helium ions through mild thermal annealing (Fig. S10 in Ref. [36]) [35]. Figure 5(g) is a summary of our discovery of strong positive correlations between $\Delta_{trig}$ and $T_N$ in layered honeycomb cobaltates. Interestingly, although the $Na_3Co_2SbO_6$ has a different space group (*C* 2/*m*) [47] compared to $Cu_3Co_2SbO_6$ (*C* 2/*c*) [31], it follows the same relationship between $\Delta_{trig}$ and $T_N$ observed in $Cu_3Co_2SbO_6$ films, as highlighted by a star in Fig. 5(g). It is strong evidence that reducing $\Delta_{trig}$ in $CoO_6$ octahedra is indeed a key to destabilizing classical ground states in layered honeycomb cobaltates.

## V. SUMMARY AND DISCUSSION

Our findings above have testified to the effectiveness of strain engineering in heterostructures to destabilize unsought long-range ordering in spin liquid candidates. While previous studies in bulk have predominantly focused on the observation of fluctuating spin-disordered phases in response to external magnetic fields [15,48-50], our approach focused on direct manipulation of the spin Hamiltonian parameters. We believe this rather unexplored research direction could allow us to observe a new side of this notoriously elusive quantum phase of matter.

For example, strain-dependent $T_N$ observed here can help to determine the effect of hopping integral for spin exchange interaction in cobalt-based honeycomb oxides. Several theoretical studies have raised questions about the effect of Co-Co direct spin-exchange interaction, which was neglected in the original calculation in high-spin $d^7$ compounds. In particular, the direct hopping integral becomes the dominant one compared to the ligand-mediated hopping when the Co-Co bond length is small enough like in $BaCo_2(AsO_4)_2$ [51,52]. This complexity of energy hierarchy in tight-binding Hamiltonian is important in Kitaev physics as the inclusion of direct hopping integral results in isotropic or XXZ-type spin Hamiltonian in honeycomb cobaltates [52-54]. In the case of thickness-controlled films, the in-plane lattice stretching is clearly shown in Fig. 3(d). This stretching leads to an elongation of the Co-Co bond length, resulting in a decrease of direct spin exchange interaction. Therefore, if this direct exchange interaction is dominant in $Cu_3Co_2SbO_6$, we would expect to see a decrease of $T_N$ in 3 u.c. $Cu_3Co_2SbO_6$ film, in opposed to our experiment in Fig. 5(c). Therefore, our observation implies that a more important factor in controlling the magnetic ground state is $\Delta_{trig}$ than direct exchange hopping at least in honeycomb cobaltates $Cu_3Co_2SbO_6$. The difference between $BaCo_2(AsO_4)_2$ and $Cu_3Co_2SbO_6$ might be attributed to different bond-length (3.11 Å for $Cu_3Co_2SbO_6$ and 2.9 Å for $BaCo_2(AsO_4)_2$) as the previous theory mentioned [52].

Our results motivate further studies on layered honeycomb cobaltate heterostructures to realize a genuine quantum spin liquid state. As several theoretical studies have pointed out, [8,27,55,56] fine-tuning of spin Hamiltonian seems to be inevitable to realize an otherwise fragile spin liquid state. However, challenges arise from the low volume of ultrathin film and/or helium-induced paramagnetic impurities, which hinder conventional magnetic susceptibility measurement and complicate the investigation of antiferromagnetic ordering in Kitaev QSL candidates. The obvious future research would be identifying substrates and buffer layers that can not only facilitate larger compressive strain but also potentially applicable to other cobaltates like $Na_3Co_2SbO_6$. Since $Na_3Co_2SbO_6$ inherently possesses a smaller $\Delta_{trig}$ than $Cu_3Co_2SbO_6$, the complete suppression of antiferromagnetic ordering can be achieved via moderate compressive strain. While searching for evidence of Kitaev QSL will be notoriously challenging in heterostructures, a few recent theoretical suggestions based on electrical measurement, spintronics as well as inelastic tunneling experiments can be applicable to our systems [57-60].

In conclusion, we have demonstrated the control of the $\Delta_{trig}$ of $Cu_3Co_2SbO_6$, a promising candidate of Kitaev QSL, and the subsequent destabilization of its classical antiferromagnetic ground state. Our lattice-engineered $Cu_3Co_2SbO_6$ system opens the pathway to tailor the spin interactions in layered honeycomb cobaltate systems, offering valuable insights into the underlying physics of Kitaev materials.

## VI. MATERIALS AND METHODS

### A. Sample preparation

The $Na_3Co_2SbO_6$ and $Cu_3Co_2SbO_6$ thin films were synthesized using pulsed laser deposition. For high-quality thin films, the O-faced ZnO [0001] substrates were annealed for 2 hours at 1100 °C in ambient pressure. The $Na_3Co_2SbO_6$ and $Cu_3Co_2SbO_6$ targets were synthesized using the solid-state reaction method with the reported recipe of polycrystalline powder [31,47]. The optimized growth conditions of $Na_3Co_2SbO_6$ were as follows: substrate temperature $T = 625$ °C, oxygen partial pressure $P = 1$ mTorr, the energy of the KrF Excimer laser ($\lambda = 248$ nm) $E = 1.1$ J / $cm^2$, the laser repetition rate is 10 Hz, and the distance between the target and substrate was set at 50 mm. The cooling process was performed under the same as grown pressure after the deposition was completed. For the XAS experiment, 40 u.c. (21 nm) thickness $Na_3Co_2SbO_6$ films were used. For $Cu_3Co_2SbO_6$, optimized growth conditions were a substrate temperature $T = 800$ °C, oxygen partial pressure $P = 10$ mTorr, the energy of the KrF Excimer laser ($\lambda = 248$ nm) $E = 1.3$ J / $cm^2$, laser repetition rate is 10 Hz, and the distance between the target and substrate was set at 50 mm. The cooling process was performed under the same as grown pressure after the deposition was completed. For the helium ion implantation, 20 u.c. (23 nm) thickness $Cu_3Co_2SbO_6$ films were used.

### B. Helium ion implantation

Helium ions were implanted into 20 u.c. $Cu_3Co_2SbO_6$ thin films using a metal ion implanter at in Korean Multi-Purpose Accelerator Complex (KOMAC). Helium ion is injected into each sample with 10 keV energy at room $T$ and vacuum environment. Each flux density of helium ion was $1.7 \times 10^{15}$ $cm^{-2}$, $2.6 \times 10^{15}$ $cm^{-2}$, and $3.6 \times 10^{15}$ $cm^{-2}$. To prevent the damage of films due to high-energy ion beam, a 48 nm-thick gold layer was deposited using thermal evaporation. For the XAS measurements, the gold layer was peeled off by Kapton tape. Details about the thickness of the gold layer and average helium ion density in the thin film are determined by Stopping and Range of Ions in Matter / Transport of Ions in Matter (SRIM/TRIM) Monte Carlo simulation (Fig. S11 in Ref. [36]).

### C. Characterization of lattice structure

To characterize the crystal structure of $Cu_3Co_2SbO_6$ thin films, high-resolution XRD data for each $Cu_3Co_2SbO_6$ thin film were collected by using D8 Advance High-Resolution X-ray Diffractometer (Bruker) with Cu $K$-$\alpha$1 wavelength. 0D Lynxeye detector is used in $\theta$-$2\theta$ scan, which has 0.01° increment with a scan speed of 0.5 sec/step. Each scan has range from 60° to 80° for thickness-dependent $Cu_3Co_2SbO_6$ thin films and from 25° to 40° for helium-implanted thin film. We used the (008) diffraction peak to determine the lattice constant $c^*$ of thickness-controlled $Cu_3Co_2SbO_6$ to avoid overlapping with ZnO substrate peak. For a detailed structural analysis, RSM data of the $Cu_3Co_2SbO_6$ (606) peak and the ZnO (114) peak were conducted on each $Cu_3Co_2SbO_6$ thin film at the 3A Hard X-ray Scattering Beamline in PLS-II of the Pohang Accelerator Laboratory.

### D. X-ray absorption spectroscopy

The XAS measurements were conducted at the 2A magnetic spectroscopy beamline in PLS-II of the Pohang Accelerator Laboratory. Those were performed in a zero-applied magnetic field, in a vacuum with a pressure lower than $2 \times 10^{-9}$ torr at room temperature, and in total electron yield mode. To prevent the charging effect, we bridged four edges of film surfaces and a copper holder with silver paint. Since the implanted helium ions can escape from the samples in high temperature [35], the sample temperature was maintained at room temperature with nitrogen gas during the vacuum baking process. We acquired XAS spectra at the Co $L$ edges and O $K$ edge with an energy resolution of 0.1 and 0.2 eV, respectively. To compare the XAS spectra of different films, they are normalized based on the total area of the isotropic spectra in the range of 770 eV to 790 eV for Co $L_3$ edge and of 785 eV to 805 eV for Co $L_2$ edge (Fig. S14 in Ref. [36]). The absolute area of XLD, $I_{|XLD|}$ is determined by integrating within the range of 775 eV to 785 eV for Co $L_3$ edge in XLD spectra. Additional XAS data for each thickness-controlled and helium ion implanted $Cu_3Co_2SbO_6$ thin films are shown in the supplemental material (Figs. S12-S15 in Ref. [36]).

### E. Crystal field multiplet calculation for Co $L$-edge XAS

The crystal field multiplet calculations were conducted using the quantum many-body script language, Quanty [40]. The user interface called Crispy is used to generate input files consisting of Slater-Condon and crystal field parameters and to visualize the simulated spectra of $I_{iso}$ and $I_{XLD}$ [41]. With $C_{3v}$ site symmetry, the trigonal field parameters $10Dq$, $D_\sigma$, and $D_\tau$ were estimated by simulating the experimental $L_3$-edge spectra. Note that both $D_{3d}$ and $C_{3v}$ have the same trigonal symmetry. The Slater-Condon parameters for electron-electron repulsion ($F^k_{dd}$, $F^k_{pd}$) and exchange ($G^k_{pd}$, $G^k_{dd}$) were scaled down by multiplying a factor of 80% to compensate for the electronic delocalization in $3d$ orbital [38]. The spin-orbit coupling parameters of $2p$ and $3d$ are evaluated as 9.745 eV and 0.066 eV, respectively. To reproduce the exact shape of each spectrum, 0.4 eV Gaussian broadening is adopted. The Lorentzian broadening for $L_2$ edge and $L_3$ edge are selected as 0.5 eV and 0.6 eV, respectively.

### F. Ellipsometry

We measured the ellipsometry parameters Ψ and Δ of $Cu_3Co_2SbO_6$ thin films on ZnO substrate by using an M-2000 ellipsometer (J. A. Woolam Co.). tanΨ is the amplitude ratio of the reflected $p$- and $s$-waves, while Δ represents the phase shift between the two waves. For all samples Ψ and Δ were obtained across an energy range from 0.74 to 6.46 eV (5900 to 52000 cm$^{-1}$) at an incident angle of 60° (80° for 3 u.c. $Cu_3Co_2SbO_6$ for adapting good sensitivity) and over a temperature range of 6 to 300 K. Each data collection lasted 200 seconds. For low-$T$ measurement, the window effect was calibrated by using a 25 nm $SiO_2$ / Si wafer. To prevent degradation of the sample implanted helium ions, we avoided baking out the chamber. The base pressure has remained below $1 \times 10^{-8}$ Torr. All samples were mounted using carbon tape to oxygen-free copper cones.

### G. Magnetic susceptibility

Field-cooled magnetization curves with respect to $T$ for $Cu_3Co_2SbO_6$, $Na_3Co_2SbO_6$, and He-implanted $Cu_3Co_2SbO_6$ thin films on ZnO substrate were measured by superconducting quantum interference device (Quantum Design). The mass and dimension of each sample were precisely investigated as 13.6 mg and 3.3 mm × 2.5 mm. We measured magnetization with the direction of ZnO (110) Also, to compensate paramagnetic signal from the substrate, the magnetization of ZnO with the same dimension, of 13.6 mg was obtained. During measurements, samples are attached to a quartz plate mounted by GE varnish. All magnetic susceptibility data were calculated by substituting the diamagnetic signal in the substrate with mass-normalization.

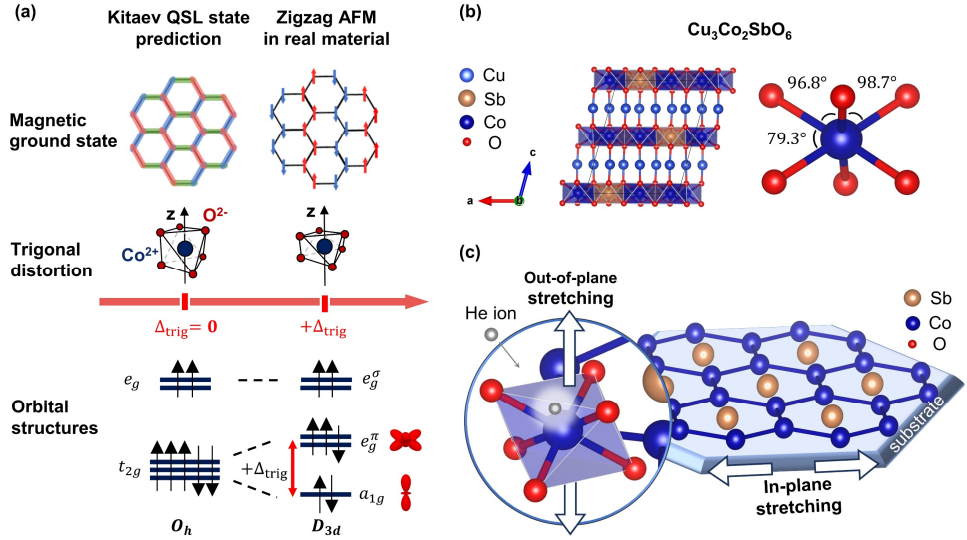

**FIG. 1.** (a) Relationship among local trigonal distortion of $CoO_6$ octahedra, orbital structures, and the magnetic ground state of $Cu_3Co_2SbO_6$. The $t_{2g}$ orbital in the $O_h$ system split into $e_g^{\pi}$ orbital and $a_{1g}$ orbital by trigonal distortion of $CoO_6$ octahedra with $D_{3d}$ symmetry. The energy difference between the $e_g^{\pi}$ and $a_{1g}$ level is defined by the $\Delta_{\text{trig}}$. According to theoretical predictions, the Kitaev QSL state could be realized with decreasing $\Delta_{\text{trig}}$ in the layered honeycomb cobaltates. (b) The crystal structure of layered honeycomb cobaltate $Cu_3Co_2SbO_6$ and O-Co-O bond angles of $CoO_6$ octahedra, indicating compressed octahedra along the out-of-plane direction [**z**-axis in Fig. 1(a)]. (c) Two ways for modulating $\Delta_{\text{trig}}$ of $CoO_6$ octahedra. While helium implantation would decrease $\Delta_{\text{trig}}$ by stretching the octahedra along the out-of-plane direction, the strain imposed on the ultrathin film through the substrate would increase $\Delta_{\text{trig}}$ by stretching the octahedra in the in-plane direction.

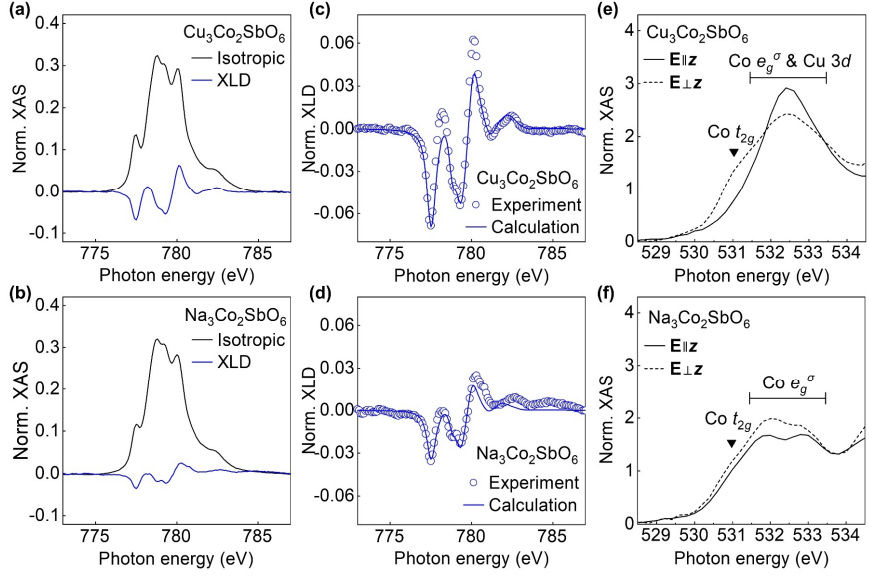

**FIG. 2.** Isotropic XAS, $I_{iso} = (2I_{\perp z} + I_{\|z})/3$, and XLD, $I_{XLD} = I_{\perp z} - I_{\|z}$, of Co $L_3$ edge in (a) $Cu_3Co_2SbO_6$ and (b) $Na_3Co_2SbO_6$. $I_{\perp z}$ ($I_{\|z}$) is normalized XAS signals with each polarization, **E** field perpendicular (parallel) to the out-of-plane direction. The experimental XLD data (open circles) and the simulation from crystal field multiplet calculation (solid lines) of (c) $Cu_3Co_2SbO_6$ and (d) $Na_3Co_2SbO_6$. Based on calculations within trigonal symmetry, the trigonal field strength is deduced to be 37.8 and 25.1 meV for $Cu_3Co_2SbO_6$, and $Na_3Co_2SbO_6$, respectively. The polarization-dependent O $K$-edge XAS for (e) $Cu_3Co_2SbO_6$ and (f) $Na_3Co_2SbO_6$. Individual peaks can be assigned to unoccupied O $2p$ orbitals hybridized with Co $3d$ $t_{2g}$ and $e_g$ orbitals. In $Cu_3Co_2SbO_6$, the additional peak near 532.5 eV is the known transition from O $1s$ to O $2p$ – Cu $3d$ hybridized state in $Cu^+$ valence systems. In both compounds, the intensity of the Co $t_{2g}$ peak is enhanced in $I_{\perp z}$, indicating dominant $e_g^\pi$ symmetry in the unoccupied state of the $Co^{2+}$ ion.

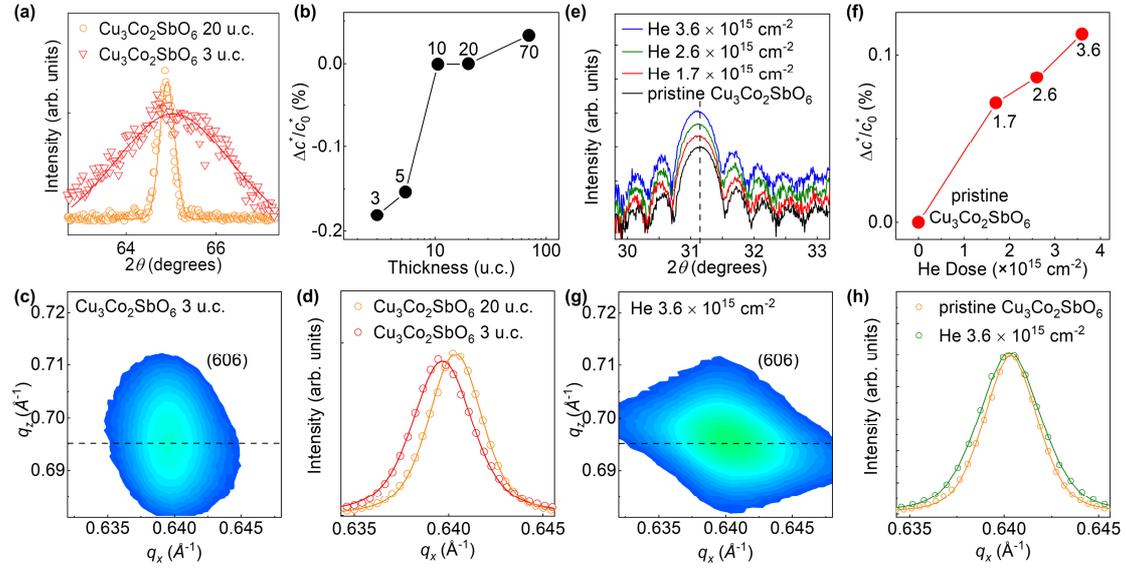

**FIG. 3.** (a) XRD $\theta$-$2\theta$ scan of near the $Cu_3Co_2SbO_6$ (008) peak of 3 u.c. and 20 u.c. film in linear scale. The tensile strain from the ZnO substrate induced expansion of the out-of-plane lattice parameter, resulting in a shift of the (008) peak of $Cu_3Co_2SbO_6$ to a higher angle. (b) The change of out-of-plane lattice parameter, $\Delta c^*/c^*_0$ as a function of thickness in $Cu_3Co_2SbO_6$ thin film. $\Delta c^*$ was calculated with respect to $c^*$ of 20 uc film, $c^*_0$. Due to the huge lattice mismatch between ZnO and $Cu_3Co_2SbO_6$, the tensile strain rapidly relaxed as the thickness of the film increased. (c) RSM and (d) $q_x$ profile for (606) plane of 3 u.c. $Cu_3Co_2SbO_6$ thin film. A smaller $q_x$ value in 3 u.c. thin film than that in 20 u.c. thick film indicates the in-plane stretching in ultrathin film (Fig. S6 in Ref. [36]). (e) XRD $\theta$-$2\theta$ scan and (f) $\Delta c^*/c^*_0$ of $Cu_3Co_2SbO_6$ film as a function of doses of helium ion. The (004) peak of $Cu_3Co_2SbO_6$ shifts to lower angles with increased dosed helium ions, indicating out-of-plane stretching of the lattice. (g) RSM image and (h) $q_x$ profile of the He $3.6 \times 10^{15}$ cm$^{-2}$ implanted $Cu_3Co_2SbO_6$ thin film (Fig. S6 in Ref. [36]). The peak from helium implanted film maintains the same $q_x$ position as that of the pristine 20 u.c. film due to its epitaxially locked in-plane lattice constants.

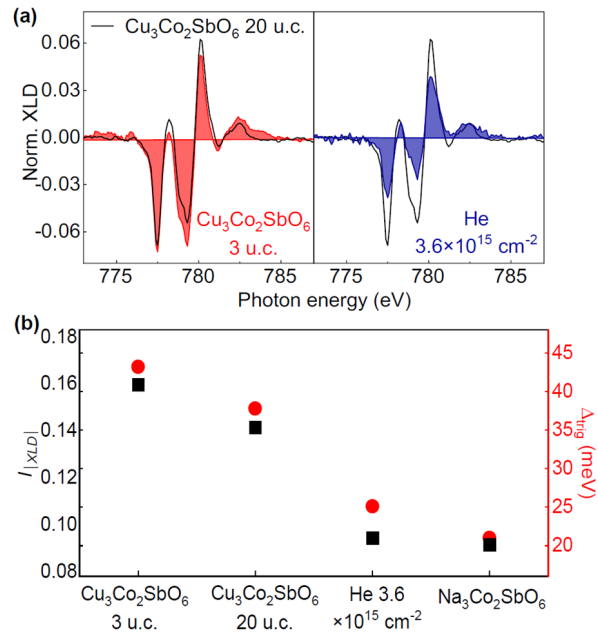

**FIG. 4.** (a) $I_{XLD}$ of Co $L_3$ edge for 3 u.c. $Cu_3Co_2SbO_6$ thin film (left) and He $3.6 \times 10^{15}$ cm$^{-2}$ implanted $Cu_3Co_2SbO_6$ thin film (right). For comparison, $I_{XLD}$ of 20 u.c. $Cu_3Co_2SbO_6$ is also shown as black lines. Films with more out-of-plane stretching showed decreased XLD signals, while films with increased in-plane stretching showed larger XLD signals. This observation clearly indicates the successful modulation of the $\Delta_{trig}$. (b) The absolute area of XLD, $I_{|XLD|}$, and $\Delta_{trig}$ deduced from the crystal field multiplet calculation of XLD spectra for 3 u.c. $Cu_3Co_2SbO_6$ film, 20 u.c. $Cu_3Co_2SbO_6$ film, He $3.6\times10^{15}$ cm$^{-2}$ implanted $Cu_3Co_2SbO_6$ film, and $Na_3Co_2SbO_6$ film.

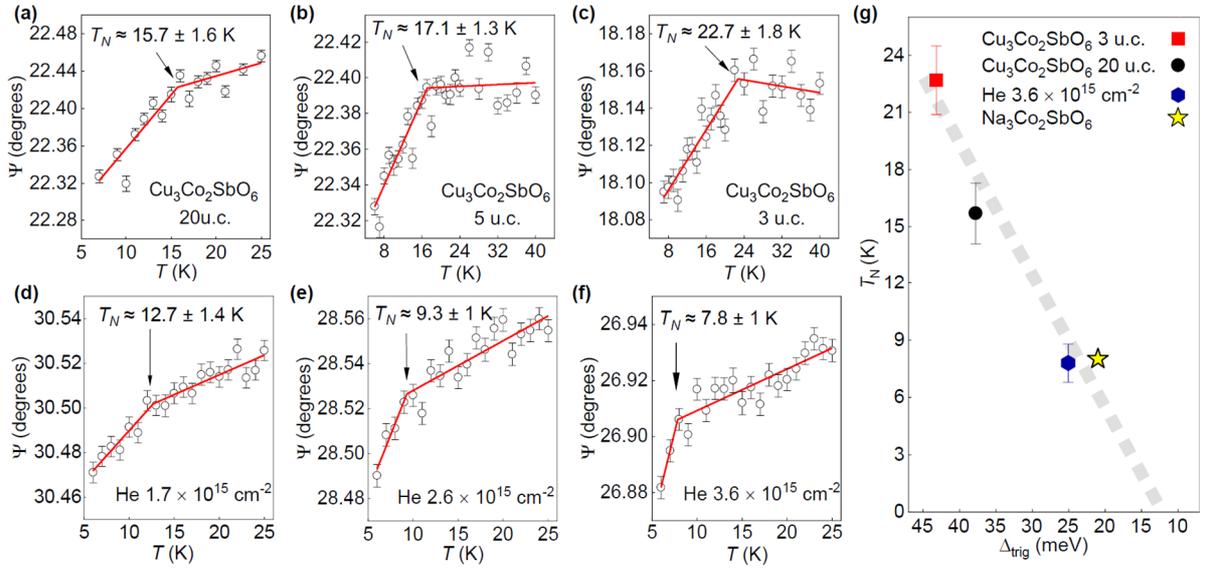

**FIG. 5.** The $T$-dependent ellipsometric parameter $\Psi$ (open circles) at exciton peak energies, which can be used to detect $T_N$ via spin-exciton coupling: The red solid lines are piecewise function fitting for (a) 20 u.c. $Cu_3Co_2SbO_6$ thin film with the $T_N \sim 15.7 \pm 1.6$ K, (b) 5 u.c. $Cu_3Co_2SbO_6$ thin film with the $T_N \sim 17.1 \pm 1.3$ K, (c) 3 u.c. $Cu_3Co_2SbO_6$ thin film with the $T_N \sim 22.7 \pm 1.8$ K, (d) He $1.7\times10^{15}$ cm$^{-2}$ implanted $Cu_3Co_2SbO_6$ with the $T_N \sim 12.7 \pm 1.4$ K, (e) He $2.6\times10^{15}$ cm$^{-2}$ implanted $Cu_3Co_2SbO_6$ with the $T_N \sim 9.3 \pm 1$ K, and (f) He $3.6\times10^{15}$ cm$^{-2}$ implanted $Cu_3Co_2SbO_6$ with the $T_N \sim 7.8 \pm 1$ K. Almost 90 % of modulation in $T_N$ is achieved by in-plane and out-of-plane stretching of $Cu_3Co_2SbO_6$ films. (g) The $\Delta_{trig}$ versus $T_N$ graph clearly visualizes strong positive correlations between those two parameters in the layered cobaltates. The grey dashed line is guided for eyes. Although $Na_3Co_2SbO_6$ has a different space group from $Cu_3Co_2SbO_6$, it follows the same relationship between $T_N$ and $\Delta_{trig}$ found in $Cu_3Co_2SbO_6$, as highlighted by a star (Fig. S9 in Ref. [36]). Therefore, the $\Delta_{trig}$ can be considered the key parameter for destabilizing classical magnetic ground state in the layered honeycomb cobaltates.

# ACKNOWLEDGMENTS

We extend our sincere gratitude to F. M. F. de Groot for his invaluable discussions. This work was mainly supported by the National Research Foundation (NRF) of Korea funded by the Ministry of Science and ICT(Grant No. NRF-2020R1C1C1008734), under the ITRC (Information Technology Research Center) support program (IITP-2023-RS-2023-00259676) supervised by the IITP (Institute for Information & Communications Technology Planning & Evaluation), and by the MSIT and PAL, Korea. Experiments at PLS-II were supported in part by MSIT and POSTECH. The exceptional assistance provided by S.Y. Park and Y.H. Kim during the PLS-II experiments is gratefully acknowledged. The preparation of helium-implanted samples was supported by the KOMAC (Korea Multi-purpose Accelerator Complex) operation fund of KAERI (Korea Atomic Energy Research Institute) by MSIT. H.R. Jeon is acknowledged for helping helium implantation experiments. The M-2000 ellipsometer (J.A. Woolam Co.) for optical measurements was supported by the IBS Center for Correlated Electron Systems, Seoul National University. Deok-Yong Cho was supported by the National Research Foundation of Korea (Grant No. 2021R1A2C1004644), funded by the Korea government (MSIT). Jung-woo Yoo was supported by the NRF of Korea (Grant No. 2021R1A2C1008431). Jong Mok Ok was supported by the Nano & Material Technology Development Program through the National Research Foundation of Korea (NRF) funded by the Ministry of Science and ICT (RS-2023-00281839).

# SUPPLEMENTAL MATERIAL

# Suppression of Antiferromagnetic Order by Strain in Honeycomb Cobaltate: Implication for Quantum Spin Liquid

Gye-Hyeon Kim,[1][*] Miju Park,[1][*] Uksam Choi,[1] Baekjune Kang,[1] Uihyeon Seo,[1] GwangCheol Ji,[2] Seunghyeon Noh,[3] Deok-Yong Cho,[4] Jung-Woo Yoo,[3] Jong Mok Ok,[2] and Changhee Sohn[1][†]

[1]Department of Physics, Ulsan National Institute of Science and Technology,
Ulsan, 44919, Republic of Korea.

[2]Department of Physics, Pusan National University, Pusan, 46241, Republic of Korea.

[3]Department of Materials Science and Engineering, Ulsan National Institute of Science and Technology,
Ulsan, 44919, Republic of Korea.

[4]Department of Physics, Jeonbuk National University, Jeonju, 54896, Republic of Korea.

# Contents

# 1. Layered structure of Co-based $d^7$ honeycomb oxides $A_3Co_2SbO_6$ (A= Na, and Cu)

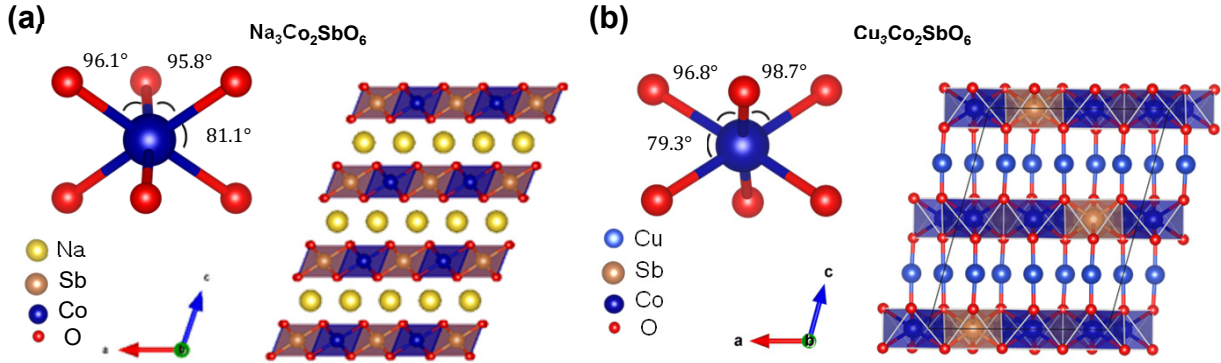

**FIG. S1.** The Co-based $d^7$ layered honeycomb oxides, (a) $Na_3Co_2SbO_6$ ($C\,2/m$ space group) and (b) $Cu_3Co_2SbO_6$ ($C\,2/c$ space group). In both cases, a Co-Sb layer and an A-site cation layer ($Na^+$ and $Cu^+$, respectively) are stacked similarly to the well-known delafossite-type structure. The Co-Sb layer has $CoO_6$ octahedra with compressive trigonal distortion.

# 2. Effect of $Sb^{5+}$ and $O^{2-}$ crystal field in $Co^{2+}$ orbitals

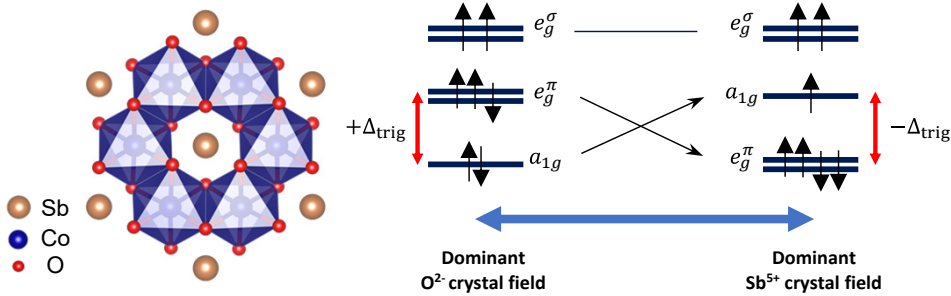

**FIG. S2.** Top view of layered honeycomb cobaltates with triangular $Sb^{5+}$ lattice (left) and orbital structures of $Co^{2+}$ ion with respect to trigonal crystal field $\Delta_{trig}$ (right). The orbital structures of $Co^{2+}$ ions are influenced by both $Sb^{5+}$ and $O^{2-}$ crystal field. Note that, the sign of the field from $Sb^{5+}$ ions is opposite to that from $O^{2-}$ ions due to its positive charge. Therefore, the sign of the total $\Delta_{trig}$ is determined by whether the crystal field from $Sb^{5+}$ or $O^{2-}$ ions is dominant.

## 3. X-ray absorption spectroscopy measurements setup

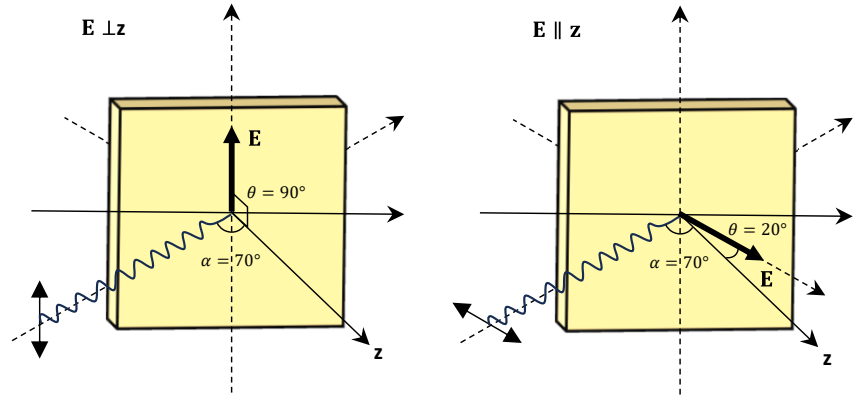

**FIG. S3.** Experimental geometry of X-ray absorption spectroscopy with surface horizontally polarized (*p*-polarized, left) and surface vertically polarized (*s*-polarized, right) incident beam. The $\theta$ is defined as the angle between the electric field vector **E** and the out-of-plane direction of the samples. The angle $\alpha$, the angle between the out-of-plane direction of samples and the incident angle of the X-ray beam, has set to 70°.

## 4. Peak assignment in O K-edge X-ray absorption spectra for $Na_3Co_2SbO_6$ and $Cu_3Co_2SbO_6$

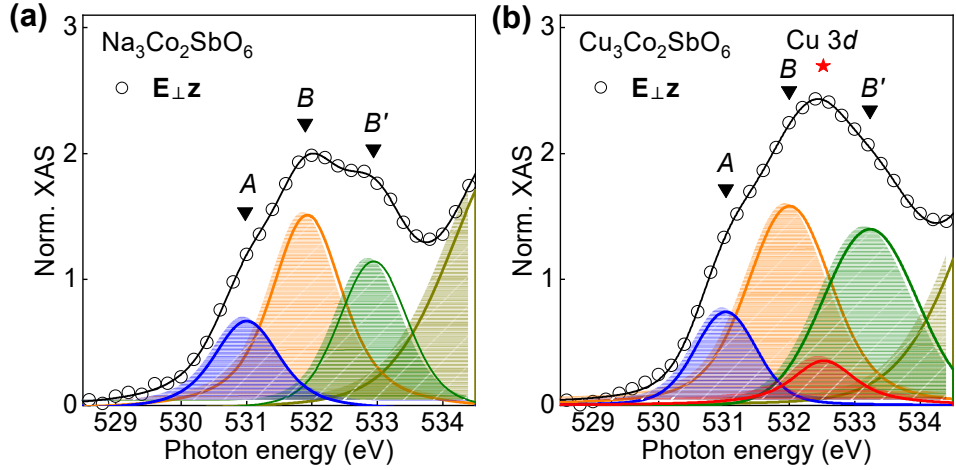

**FIG. S4.** (a) $I_{\perp z}$ of O K-edge X-ray absorption spectroscopy in $Na_3Co_2SbO_6$. Three peaks labeled A, B, and B' are associated with the Co 3d orbitals. Peak A is assigned as the electron transition from O 1s to O 2p orbitals hybridized Co $t_{2g}$ orbitals (final state of $^3A_2$) [44]. Both peak B and B' are from O 2p orbitals hybridized Co $e_g^\sigma$ orbitals with final states of $^3T_2$ and $^3T_1$, respectively [44]. (b) $I_{\perp z}$ of O K-edge X-ray absorption spectroscopy in $Cu_3Co_2SbO_6$. Similar peak features of A, B, and B' are also shown at almost identical photon energy. Note that Cu 3d peak at 532.5 eV is related to the O 2p states hybridized with Cu 3d orbitals in $Cu^+$ ions [45,46].

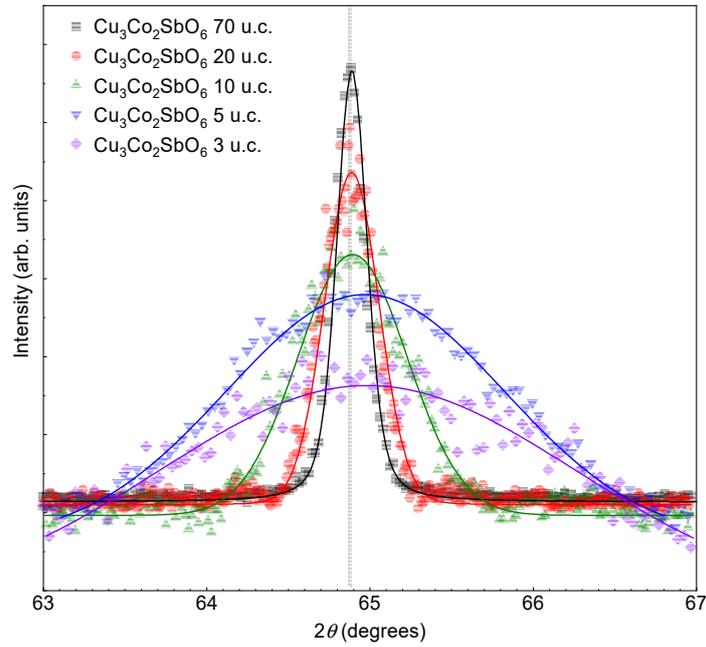

**FIG. S5.** $\theta$-$2\theta$ (00$l$) scan of ZnO (008) plane for thickness-dependent $Cu_3Co_2SbO_6$ thin films (dot line) with fitting results (solid line) in linear scale. The peak position for bulk-like 70 unit-cell (u.c.) thick $Cu_3Co_2SbO_6$ thin film is marked as a grey solid line. The (008) peak of $Cu_3Co_2SbO_6$ films are gradually shifted to a higher angle from 70 u.c. to 3 u.c. due to the tensile strain from ZnO substrate. This shift indicates the contraction in the out-of-plane lattice constant $c^*$ of thickness-controlled $Cu_3Co_2SbO_6$.

## 5. Detailed structure analyses of ultrathin and helium implanted $Cu_3Co_2SbO_6$ films

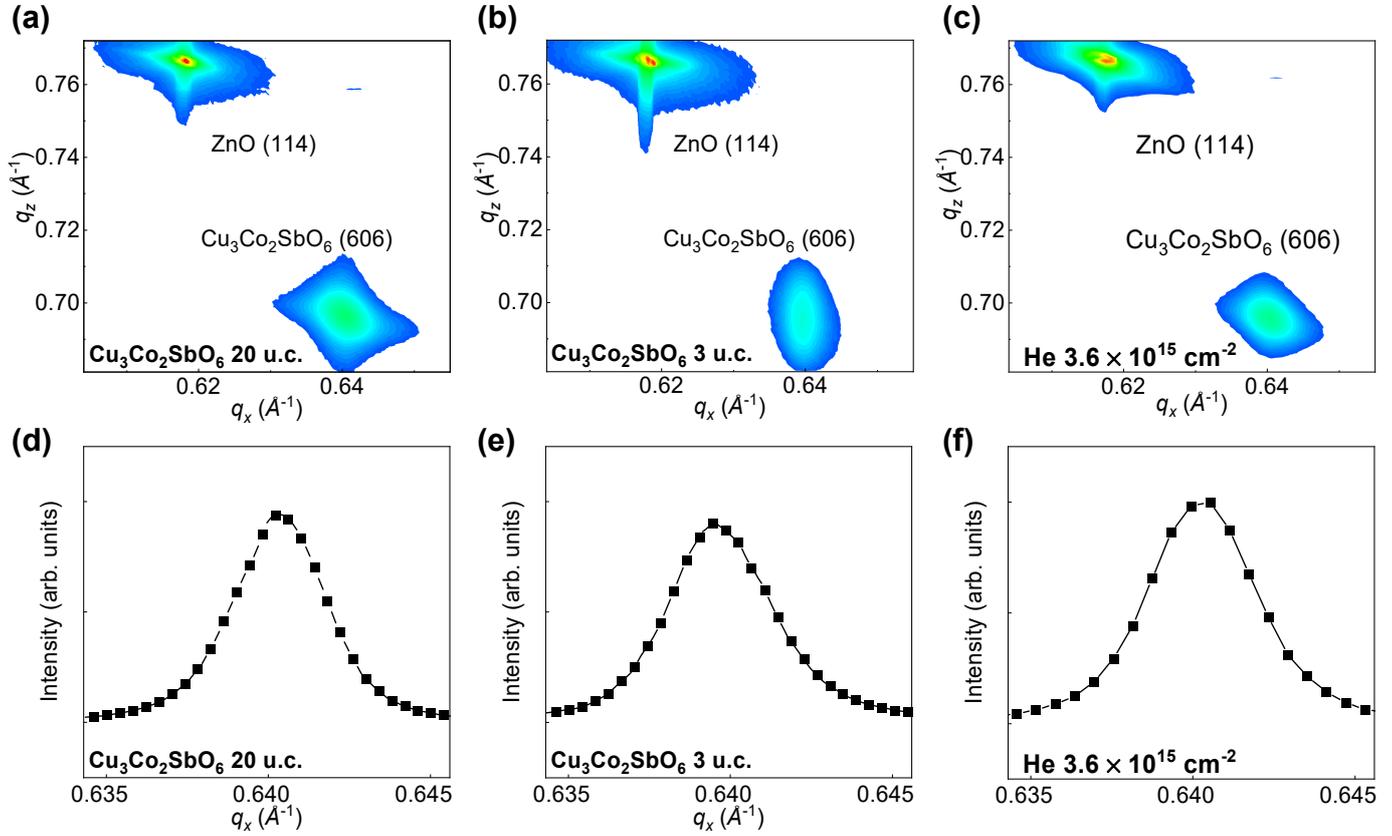

**FIG. S6.** Reciprocal space mapping of ZnO (114) plane for (a) bulk-like 20 unit-cell (u.c.), (b) ultrathin 3 u.c., and (c) He $3.6 \times 10^{15}$ cm$^{-2}$ $Cu_3Co_2SbO_6$ thin films. The (606) peak of $Cu_3Co_2SbO_6$ is shown in all thin films. The $q_x$ profiles of (d) bulk-like 20 u.c., (e) ultrathin 3 u.c., and (f) He $3.6 \times 10^{15}$ cm$^{-2}$ $Cu_3Co_2SbO_6$ thin films are extracted from the reciprocal space mapping data.

## 6. Crystal field multiplet simulation for Co $L_3$-edge X-ray linear dichroism

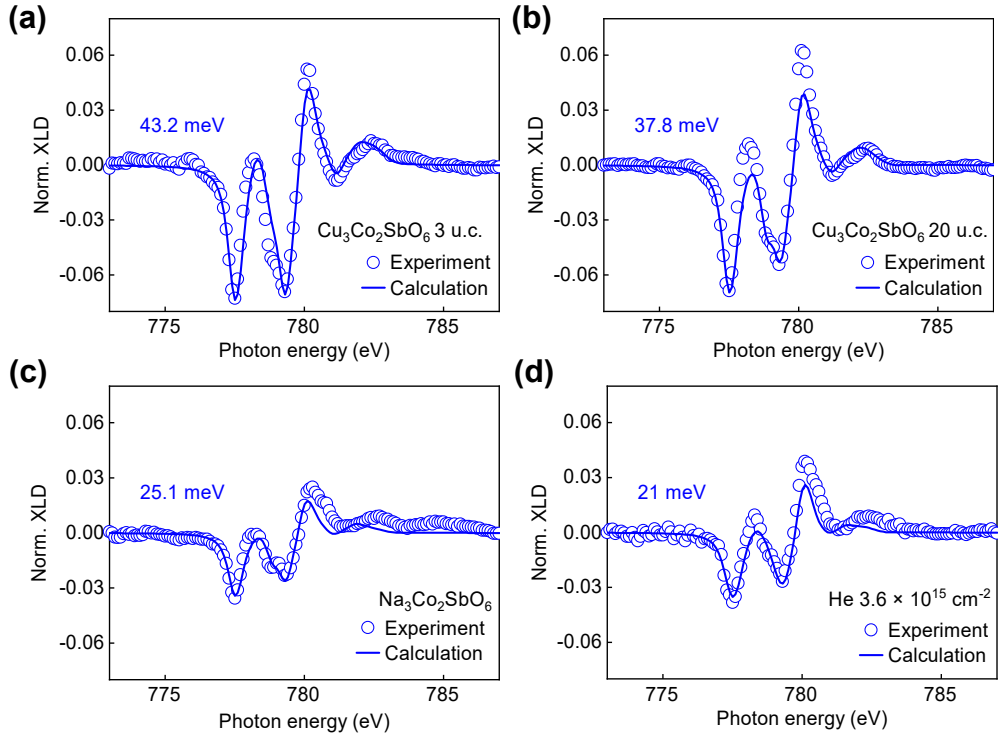

**FIG. S7.** Experimental X-ray linear dichroism data (open circle) and the simulation from crystal field multiplet calculation (solid line) of (a) 3 unit-cell (u.c.) $Cu_3Co_2SbO_6$ film, (b) 20 u.c. $Cu_3Co_2SbO_6$ film, (c) $Na_3Co_2SbO_6$ film, and (d) He $3.6 \times 10^{15}$ cm$^{-2}$ implanted $Cu_3Co_2SbO_6$ film. Within the trigonal symmetry calculation, the magnitude of the trigonal crystal field $\Delta_{trig}$ is deduced to be 43.2 meV for 3 u.c. $Cu_3Co_2SbO_6$ film, 37.8 meV for 20 u.c. $Cu_3Co_2SbO_6$, 25.1 meV for $Na_3Co_2SbO_6$, and 21 meV for He $3.6 \times 10^{15}$ cm$^{-2}$ implanted $Cu_3Co_2SbO_6$ film.

# 7. Optical spectra, excitonic transition, and spin-exciton coupling in $Cu_3Co_2SbO_6$ films

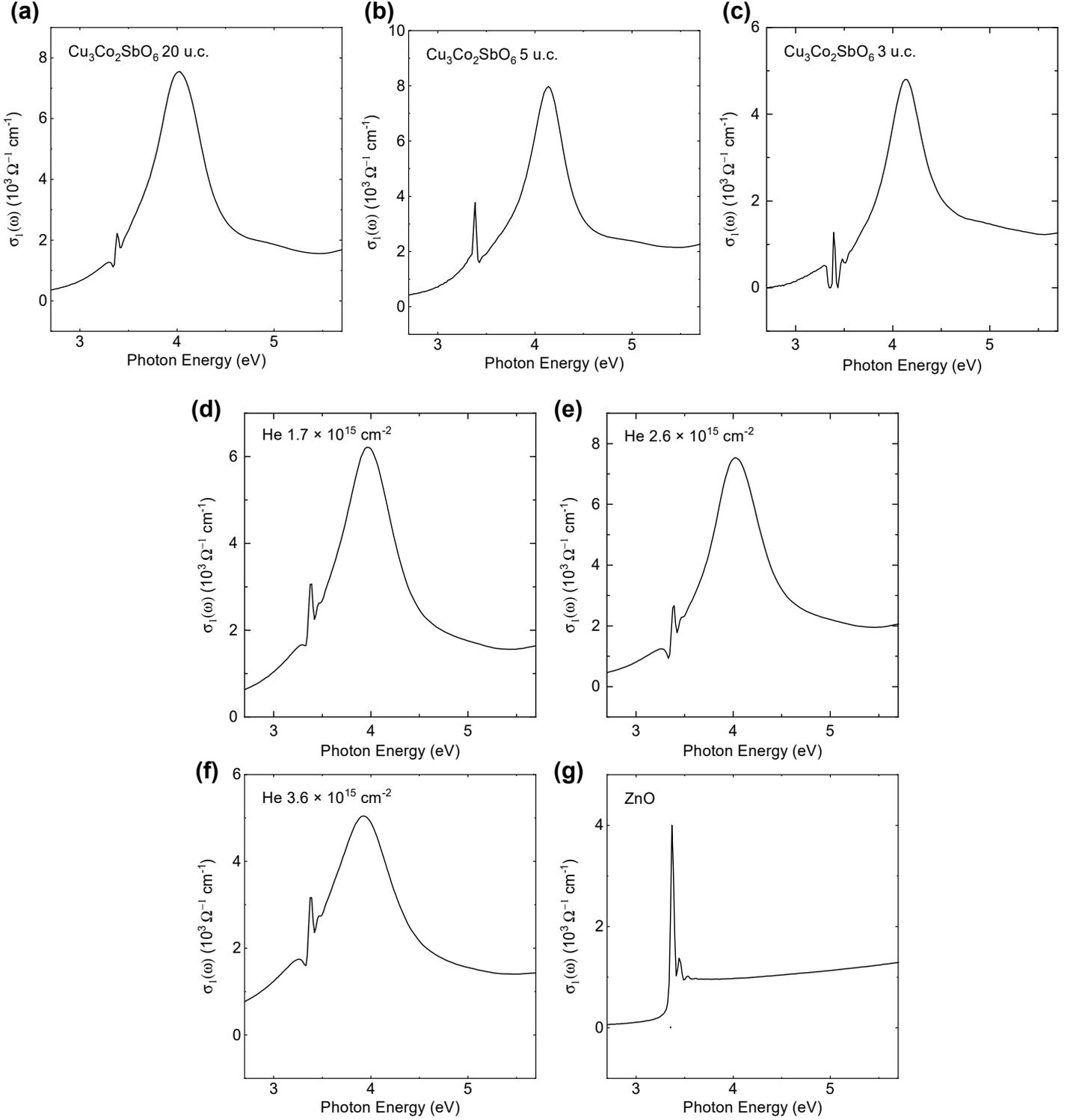

**FIG. S8.** Optical conductivity $\sigma_1(\omega)$ at 7 K of (a) 20 unit-cell (u.c.) $Cu_3Co_2SbO_6$ film, (b) 5 u.c. $Cu_3Co_2SbO_6$ film, (c) 3 u.c. $Cu_3Co_2SbO_6$ film, (d) He $1.7 \times 10^{15}$ cm$^{-2}$ implanted $Cu_3Co_2SbO_6$ film, (e) He $2.6 \times 10^{15}$ cm$^{-2}$ implanted $Cu_3Co_2SbO_6$ film, (f) He $3.6 \times 10^{15}$ cm$^{-2}$ implanted $Cu_3Co_2SbO_6$ film and (g) ZnO substrate. All the strong excitonic excitation, attributed to the Cu 3$d$ → Co 3$d$ $e_g$ transition, are located near 4 eV. A sharp anomaly near 3.35 eV in optical spectra of $Cu_3Co_2SbO_6$ films is an artifact from sharp excitations in ZnO substrates.

## 8. Magnetic susceptibility of $Na_3Co_2SbO_6$ and $Cu_3Co_2SbO_6$ thin films

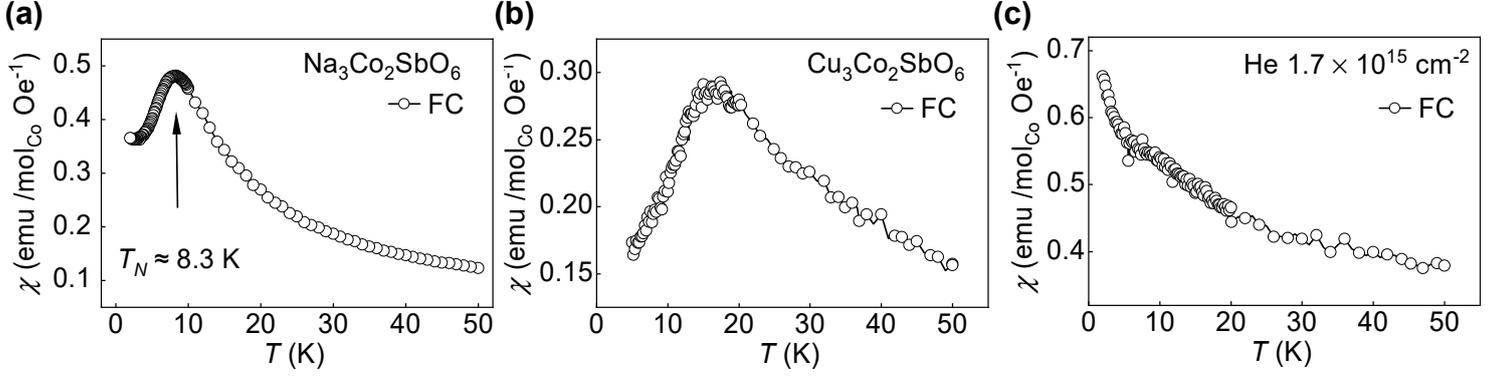

**FIG. S9.** Magnetic susceptibility $\chi(T)$ of (a) $Na_3Co_2SbO_6$ film, (b) pristine $Cu_3Co_2SbO_6$ film, and (c) He $1.7 \times 10^{15}$ cm$^{-2}$ implanted $Cu_3Co_2SbO_6$ film. All data are estimated from 0.1 T Field-cooled magnetization curves. For $Na_3Co_2SbO_6$ film and pristine $Cu_3Co_2SbO_6$ film, clear antiferromagnetic transitions are shown at about 8 K, and 16 K, respectively. It is consistent with previous studies on the bulk $Na_3Co_2SbO_6$ [44] and $Cu_3Co_2SbO_6$ [31]. However, helium implanted $Cu_3Co_2SbO_6$ exhibits additional paramagnetic signal at low $T$, indicating formation of paramagnetic impurity during helium ion implantation. It significantly hinders the precise determination of $T_N$ in helium implanted $Cu_3Co_2SbO_6$ by using conventional susceptibility measurement.

## 9. Reversal effect of helium ion implantation in crystal structure and $T_N$ of $Cu_3Co_2SbO_6$

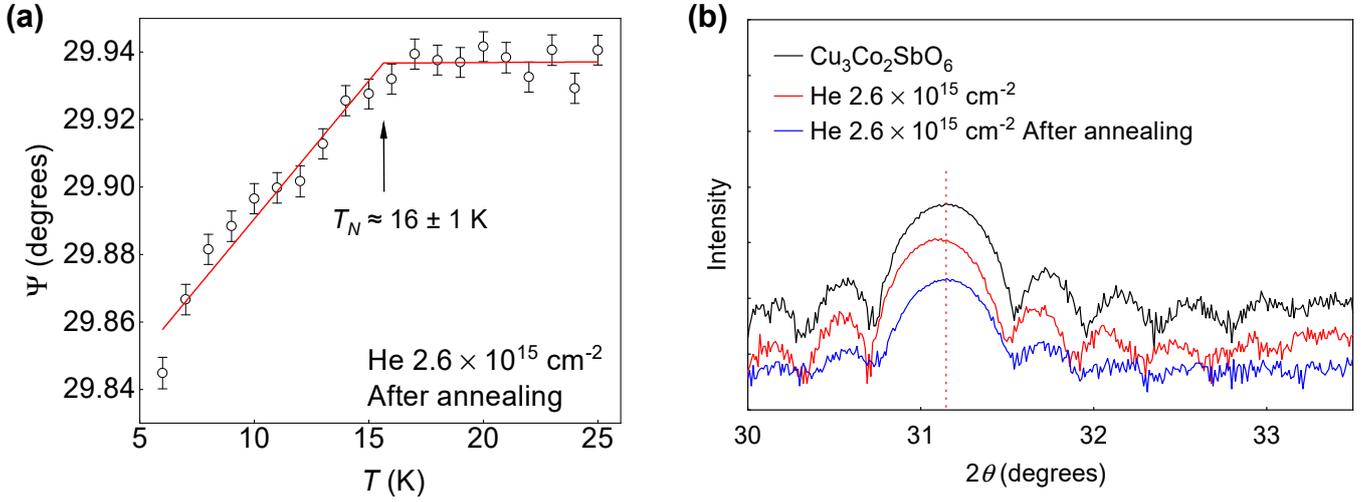

**FIG. S10.** (a) $T$-dependent ellipsometry parameter $\Psi$, where $\tan\Psi$ correspond to amplitude ratio between reflected $p$- and $s$-polarized light, and (b) X-ray diffraction of helium implanted $Cu_3Co_2SbO_6$ thin film before and after thermal annealing to remove helium ions. Helium implanted $Cu_3Co_2SbO_6$ film was annealed at 250 °C, 3 hours in ambient pressure. Both data show reversible effect of helium ion implantation via post-annealing. In (a), the $T_N$ of 9.3 K before the annealing [Fig. 5(e) in the main manuscript] is returned to 16 K like in pristine $Cu_3Co_2SbO_6$ film. In (b), the shifted (004) peak of the helium implanted $Cu_3Co_2SbO_6$ is recovered to the original position of pristine $Cu_3Co_2SbO_6$ film. Furthermore, the retained thickness fringes demonstrate the high quality of the film, with no indications of damage resulting from the implantation and removal of helium ions.

## 10. SRIM/TRIM Monte Carlo simulation for helium ion implantation

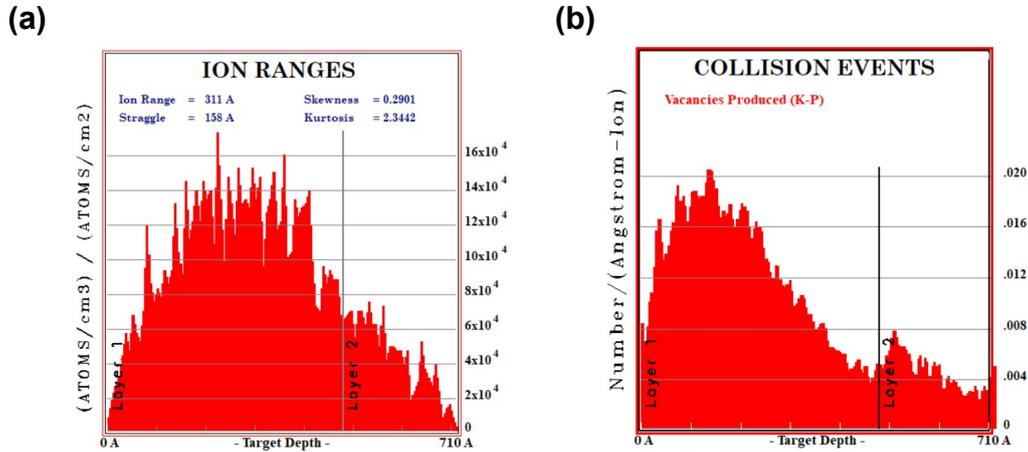

**FIG. S11.** Stopping and Range of Ions in Matter / Transport of Ions in Matter (SRIM/TRIM) Monte Carlo Simulation for helium ion implantation. (a) The spatial distribution of helium ions within a thin film composed of $Cu_3Co_2SbO_6$, with a gold layer deposited on top, when helium ions with energy of 10 keV are incident upon it. (b) The generation of vacancies resulting from helium ions collisions within the gold coated $Cu_3Co_2SbO_6$ thin film. Layer 1 is the gold layer with a thickness of 48 nm, while Layer 2 is the $Cu_3Co_2SbO_6$ layer with a thickness of 23 nm. It is noteworthy that we carefully determined the thickness of the gold layer through simulations to ensure that helium ions incident upon the $Cu_3Co_2SbO_6$ thin film do not cause substantial damage [34]. Given the linear distribution of helium ions within the $Cu_3Co_2SbO_6$ thin film, we computed the density of helium ions per unit volume in Layer 2 as half of the density value at the interface between gold and $Cu_3Co_2SbO_6$ thin film, multiplied by the incident fluence.

## 11. Full data of X-ray absorption spectroscopy measurements

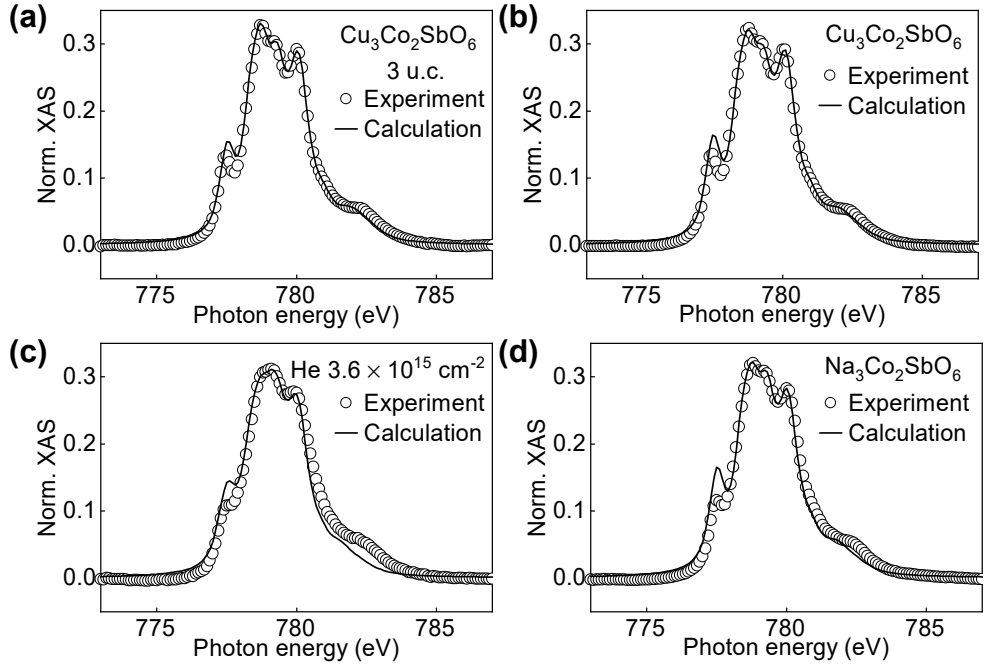

**FIG. S12.** Experimental and simulated isotropic Co $L_3$-edge spectra of (a) 3 unit-cell $Cu_3Co_2SbO_6$ film, (b) $Cu_3Co_2SbO_6$ film, (c) He $3.6 \times 10^{15}$ cm$^{-2}$ implanted $Cu_3Co_2SbO_6$ film, and (d) $Na_3Co_2SbO_6$ films.

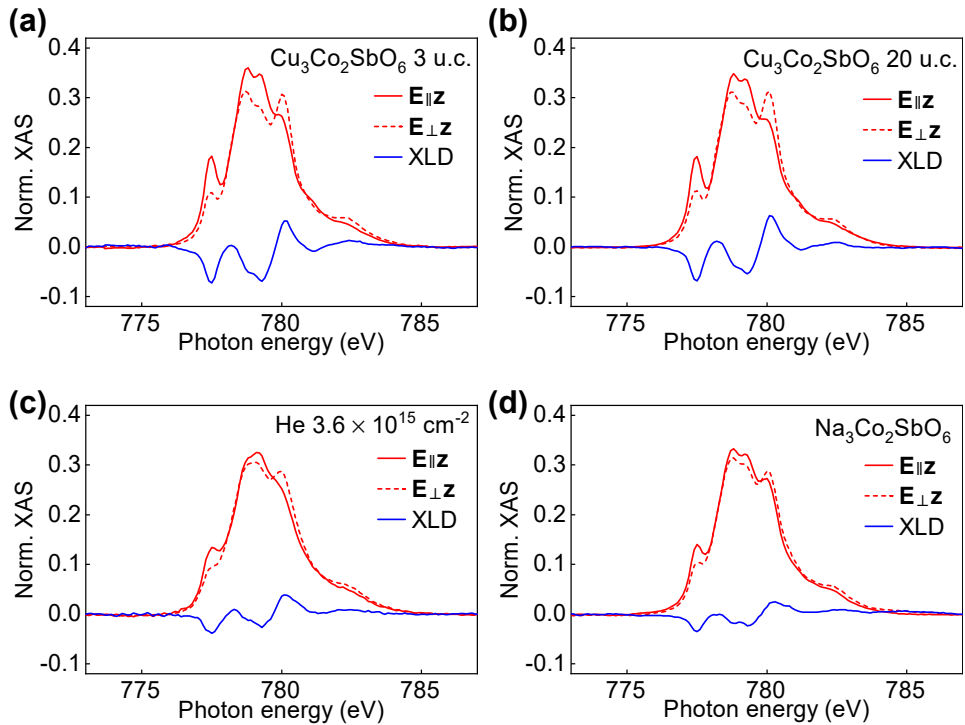

**FIG. S13.** The normalized polarized spectra and X-ray linear dichroism of (a) 3 unit-cell $Cu_3Co_2SbO_6$ film, (b) pristine 20 unit-cell $Cu_3Co_2SbO_6$ film, (c) He $3.6 \times 10^{15}$ cm$^{-2}$ implanted $Cu_3Co_2SbO_6$ film, and (d) $Na_3Co_2SbO_6$ film.

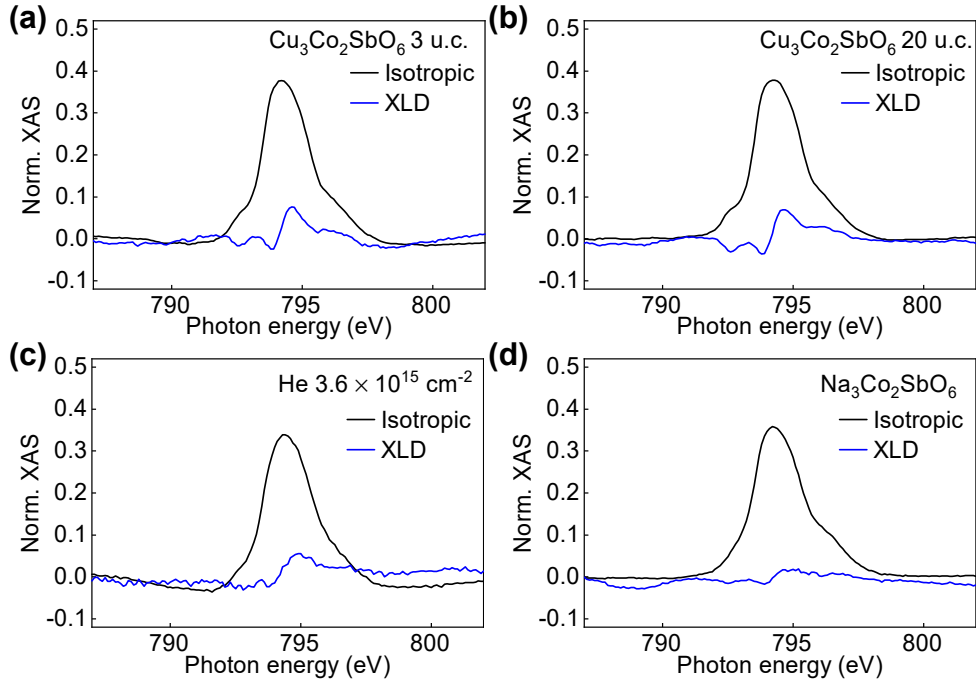

**FIG. S14.** The normalized Co $L_2$-edge isotropic and XLD spectra of (a) 3 unit-cell $Cu_3Co_2SbO_6$ film, (b) pristine 20 unit-cell $Cu_3Co_2SbO_6$ film, (c) He $3.6 \times 10^{15}$ cm$^{-2}$ implanted $Cu_3Co_2SbO_6$ film, and (d) $Na_3Co_2SbO_6$ film. The Co $L_2$-edge spectra exhibit a similar trend to the Co $L_3$-edge spectra.

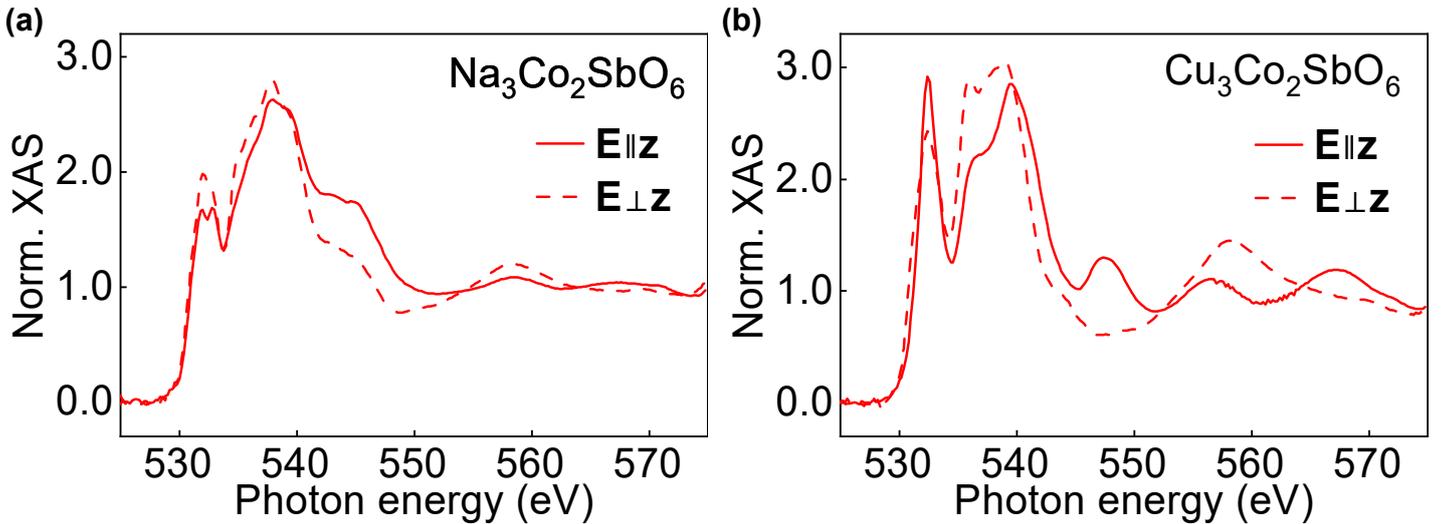

**FIG. S15.** The normalized O $K$-edge spectra of (a) $Na_3Co_2SbO_6$ film and (b) $Cu_3Co_2SbO_6$ film from 525 eV to 575 eV. It could be assumed the complicated hybridization due to the unoccupied state of the O $2p$ orbital is in high energy [42]. These O $K$-edge spectra were normalized at 545 eV–575 eV.